\documentclass[twocolumn,
superscriptaddress,
prl,
showpacs,
preprintnumbers,
amsmath,amssymb]{revtex4}

\usepackage{graphicx,epsf,epsfig,ulem}
\usepackage{epsfig}
\usepackage{bm}
\usepackage{subfigure}
\usepackage{color}
\usepackage{graphicx}

\begin{document}

\title{Mapping donor electron wavefunction deformations at sub-Bohr orbit resolution}

\author{Seung H. Park}
\affiliation{Network for Computational Nanotechnology, Purdue University, West Lafayette, IN 47907, USA}

\author{Rajib Rahman}
\affiliation{Network for Computational Nanotechnology, Purdue University, West Lafayette, IN 47907, USA}

\author{Gerhard Klimeck}
\affiliation{Network for Computational Nanotechnology, Purdue University, West Lafayette, IN 47907, USA} 

\author{Lloyd C. L. Hollenberg}
\affiliation{Center for Quantum Computer Technology, School of Physics, University of Melbourne, VIC 3010, Australia}

\date{\today} 

\begin{abstract} 
Quantum wavefunction engineering of dopant-based Si nanostructures reveals new physics in the solid-state, and is expected to play a vital role in future nanoelectronics. Central to any fundamental understanding or application is the ability to accurately characterize the deformation of the electron wavefunctions in these atom-based structures through electric and magnetic field control. We present a method for mapping the subtle changes that occur in the electron wavefunction through the measurement of the hyperfine tensor probed by $^{29}$Si impurities. We calculate Stark parameters for six shells around the donor. Our results show that detecting the donor electron wavefunction deformation is possible with resolution at the sub-Bohr radius level.
\end{abstract} 


\maketitle

The exponential miniaturization of semiconductor technology over the past 50 years has ushered in an era of nanoscale quantum electronics. At near atomic dimensions, conventional device operations are strongly affected by quantum phenomena in the solid-state {\cite{Rogge.NaturePhysics.2008}}. To ensure continued progress in semiconductor electronics, and indeed in the drive for new quantum nanoelectronic devices, the inherently quantum aspects of such systems need to be understood and even incorporated into device functionality. The possibility of harnessing quantum phenomena in devices has produced revolutionary ways of performing computing, as exemplified by the rapidly developing fields of quantum computing and spintronics {\cite{Sarma.ModernPhys.2004}}. One central concept of quantum electronics is the ability to induce controlled deformations of a specific donor-bound electron wavefunction by external electric and magnetic fields. Accessing the details of such wavefunction engineering is critical to understanding and developing new devices and applications. However, until now there has been no way of quantifying the nature of such wavefunction distortions beyond indirect means {\cite{Rogge.NaturePhysics.2008}}.

In this letter, we propose an electron-nuclear double resonance (ENDOR) experiment to directly measure the gate induced Stark shift of the donor electron hyperfine tensor at specific lattice sites near the donor site (Figure 1). $^{29}$Si atoms distributed randomly in the lattice provide a direct nuclear spin probe of the donor electron wavefunction within the Bohr orbit region. Our atomistic tight-binding simulations for lattice regions involving over a million atoms show that this technique provides a spatial map of the bound donor electron response to a controlling gate field to sub-Bohr orbit resolution, with excellent correlation to the deformed electronic wavefunction, and confirm the feasibility of detecting such field induced hyperfine resonance shifts. The technique also has wide applicability as it can in principle be extended to map out the electric field response of wavefunctions in single electron Si quantum dots, quantum wells or other nanostructures. The ability to map single electron wavefunction distortions may have far reaching consequences for many current and future quantum nanoelectronic applications. 

\begin{figure}[htbp]
\center
\epsfxsize=2.8in
\epsfbox{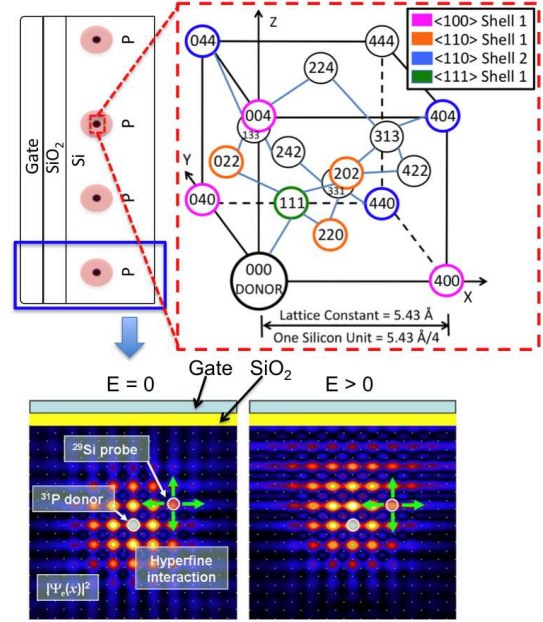}
\caption{Schematic of the technique. Top row: A series of donors in Si under a gate. Inset shows classification of the sub-Bohr radii region into shells based on symmetry and distance. Bottom row: Probing the field-induced distortions of the donor wavefunction by a ${}^{29} \textrm{Si}$ atom using hyperfine interaction.} 
\end{figure}

Silicon-based quantum nanoelectronic systems benefit from long spin coherence times and the expertise of the semiconductor industry in scalable system design and manufacture. As a result there are a number of key proposals for quantum computing devices, including substitutional donors {\cite{Kane.nature.1998, Vrijen.pra.2000, Hollenberg.prb.2004.1, Hollenberg.prb.2006, Calderon.prl.2004}}, gate-confined 2DEGs {\cite{Loss.pra.1998}}, and Si quantum wells {\cite{Eriksson.NaturePhysics.2007}}. Advances in single atom {\cite{Schofield.prl.2003}} and ion implantation {\cite{Jamieson.apl.2005}} technologies may enable repeatable fabrication of dopant-based nanostructures. Some recent structures include a single gated donor in a FinFET {\cite{Rogge.NaturePhysics.2008}}, a gated two donor charge qubit {\cite{Andresen.Nanoletter.2007}}, a 2D gated donor layer {\cite{Ruess.prb.2007}}, and an 1D metallic wire of donors {\cite{Ruess.prb.2007_1}}. The wavefunctions of such donor based nanostructures vary considerably from their bulk counterparts, yet are critical to device operation. A direct map of the wavefunctions and their electric field response will be important in novel quantum device design and engineering.   

The method we describe here uses the hyperfine interaction between a donor bound electron spin and a nuclear spin of a ${}^{29} \textrm{Si}$ isotope in a lattice of spinless ${}^{28} \textrm{Si}$ atoms, similar to the method used by Hale and Mieher {\cite{Hale.prb.1969}}, but critically we include and analyse the effect of a controlling field deforming the donor wavefunction. Although current technology limits the substitution of a ${}^{28} \textrm{Si}$ atom by a ${}^{29} \textrm{Si}$ atom at a specific point in the lattice, it is nevertheless possible to prepare ensemble device samples (Fig. 1) with  ${}^{29} \textrm{Si}$ atoms distributed randomly around a gated donor. The hyperfine interaction between a donor electron spin $S$ and a ${}^{29} \textrm{Si}$ nuclear spin $I$ is $H = \vec{I} \cdot A \cdot \vec{S}$. Taking the origin at the $^{29}$Si nucleus, the $A$ tensor is 

\begin{equation} \label{eq:tensor_eq} 
A_{ij} = \gamma_I \gamma_S \hbar^2 \left( {8\pi \over 3} |\Psi(0)|^2 + \langle\Psi | {3 r_i r_j - r^2\delta_{ij}\over r^5} | \Psi \rangle \right),
\end{equation}
where $\gamma_I$ and $\gamma_S$ are the nuclear and electronic gyromagnetic ratio respectively, and $r_{i,j}=(x,y,z)$. The first term in (1) is the Fermi contact hyperfine interaction, denoted here as $\beta$, and is directly proportional to the electronic probability density at the ${}^{29} \textrm{Si}$ site. The second term represents the magnetic dipolar or anisotropic hyperfine interaction between the two spins, denoted as $B_{ij}$. This dipolar term can also contribute to the ENDOR resonance energies providing a further measure of the distribution of the donor electron wave function about the probe site. 

ENDOR measurements were performed, first by Feher {\cite{Feher.prb.1959}} and later by Hale and Mieher {\cite{Hale.prb.1969}}, to study parts of the ground state wavefunction of a donor close to the nucleus, resolving as many as 20 shells (Fig 1) {\cite{Hale.prb.1969}}. Several theoretical models have calculated the hyperfine tensors of a few shells with semi-quantitative agreement with the experiments {\cite{Hale.prb.1971, Ivey.prl.1972}}. A recent ab-initio DFT study was able to calculate very accurately the tensor components of a few shells in the vicinity of the donor nucleus {\cite{Overhof.prl.2004}}. Changes in the Fermi contact hyperfine constants under a uni-directionally applied stress were also measured {\cite{Hale.prb.1970}}. It was shown {\cite{Witzel.prb.2007}} that inclusion of the anisotropic hyperfine interaction in spin coherence time calculations provides remarkable agreement between theory and recent measurements. The only work on the Stark shift of the hyperfine tensors to date computed the Fermi contact coupling for 3 lattice sites 
{\cite{Debernardi.prb.2006}}. 

The single donor wavefunctions subjected to constant electric fields were computed using an atomistic semi-empirical tight-binding (TB) model involving a 20 orbital per atom basis of $sp^3d^5s*$ (spin) orbitals with nearest neigbour interactions. 
The donor was modeled by a Coulomb potential with a cut-off value $U_0$ at the donor site. The magnitude of $U_0$ was adjusted to obtain the experimental binding energy of Si:P. The total Hamiltonian was diagonalized by a parallel Lanczos eigensolver to obtain the low lying impurity wavefunctions, which were then used to evaluate the hyperfine tensors using equation (1). The TB method used here is embeded in the Nanoelectronic Modeling Tool (NEMO-3D) {\cite{Klimeck.cmes.2002}}, and has been successfully applied to compute Stark shift of the donor contact hyperfine coupling {\cite{Rahman.prl.2007}} in excellent agreement with experiments {\cite{Bradbury.prl.2006}} and with momentum space methods {\cite{Wellard.prb.2005}}. The method was also used to interpret single donor transport experiments in FinFETs {\cite{Rogge.NaturePhysics.2008}}.   

Two major issues need to be addressed for practical implementation of the experiment. First, it is necessary to associate each hyperfine resonance peak with a  ${}^{29} \textrm{Si}$ lattice site. Second, electrostatic gates may give rise to inhomogeneous electric fields in the lattice, subjecting each donor to a different E-field, and may limit the distinguishability of signals corresponding to a ${}^{29} \textrm{Si}$ site.

The first issue has been resolved in previous works by classifying the lattice sites according to their symmetry and distance from the donor. For example, the lattice sites in the [100] equivalent directions from the donor are grouped into a different symmetry class as opposed to the points in the [110] or [111] directions. At $E=0$, all the points equidistant from the donor are responsible for a single hyperfine peak, and can be grouped as a shell for ease of identification. At a non-zero E-field, these points are no longer equivalent, giving rise to multiple resonance peaks. As an example, there are 6 points one lattice constant $a_0$ away from the donor along the [100] equivalent directions. If an E-field directed along [010] is applied, the four points (with lattice sites $(\pm a_0,0,0)$ and $(0,0,\pm a_0)$) lying in a plane perpendicular to the field are still equivalent, and produce a single resonance resonance peak. Since the site $(0,a_0,0)$ is at a different potential than $(0,-a_0,0)$, two separate peaks are observed. Overall, the six-fold degenerate hyperfine peak splits into 3 components (Fig 2a).

\begin{figure}[htbp]
\center
\epsfxsize=3.2in
\epsfbox{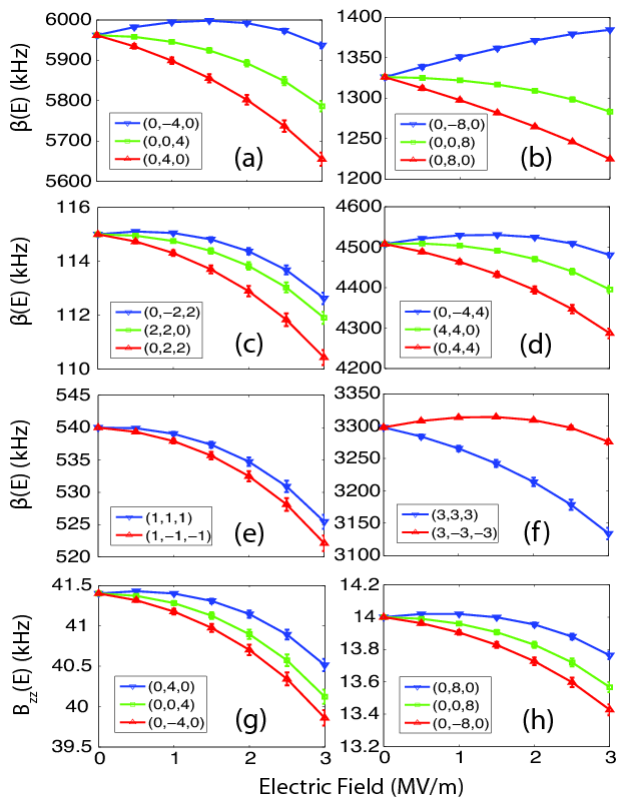}
\caption{Relative change in the contact hyperfine coupling $\beta$ as a function of electric field for two groups (shells) of points along $\langle 100 \rangle$ (a, b), $\lbrace 110 \rbrace$ (c, d), and $\langle 111 \rangle$ (e, f). The coordinates of the lattice sites are in units of $a_0/4$, where $a_0=0.543$ nm. Stark shift of a dipolar component $B_{zz}$ is also shown in (g) and (h) for two shells. The error bars are due to an uncertainty of 0.1 MV/m in the E-field.} 
\end{figure} 

To minimize the effect of inhomogeneous E-fields, it is usually advantageous to introduce the donors by ion implantation rather than by bulk doping. Ion implantation at several 100 keVs with typical doses of $10^{11}$ $\textrm{cm}^{-2}$ can ensure a rapidly decaying Gaussian depth distribution with a sharply peaked mean depth \cite{Bradbury.prl.2006, Schenkel.apl.2006}. In fact, the problem of inhomogeneous E-fields were dealt with in this manner in Refs \cite{Bradbury.prl.2006}, which successfully measured the Stark shifted hyperfine interaction between the donor electronic and nuclear spins. Simulations of the ion implantation process can also yield an estimate of the typical uncertainty in donor depths, which can be incorporated in the data analysis. In addition, clever gate designs can be utilized that combine the advantages of fairly uniform E-fields generated from a parallel plate capacitor-like structure, while making sure that the microwave radiation from the spin excitations are not completely shielded. We have also incorporated reasonable uncertainty in the E-field in our analysis to predict suitable field regimes for measurements. All these measures can help mitigate the effect of inhomogeneous fields.

\begin{center}
\begin{table} [htbp]
\caption{Qudratic ($\eta_2$) and linear ($\eta_1$) Stark coefficients for the tensor components of 6 shells around the donor in units of $\textrm10^{-3}$ $\textrm{m}^{2}/\textrm{MV}^{2}$ for $\eta_2$ and $10^{-3}$ m/MV for $\eta_1$.}
\label{tb:tablename}

{
\begin{tabular}{c cc cc c cc cc}
\hline
\hline

&
\multicolumn{2}{c}{$\beta$}
&
\multicolumn{2}{c} {$B_{zz}$} 
&

&
\multicolumn{2}{c}{$\beta$}
&
\multicolumn{2}{c} {$B_{zz}$} 

\\
Site & $\eta_2$ & $\eta_1$ & $\eta_2$ & $\eta_1$ & Site & $\eta_2$ & $\eta_1$ & $\eta_2$ & $\eta_1$\\

\hline 
(0 0 4) & -3.8 & 1.7 & -3.9 & 1.8 & (0 0 8) & -4.1 & 1.8 & -3.9 & 1.8
\\   

(0 4 0) &-3.5 & -6.4 & -3.8 & 4.5 & (0 8 0) & -2.2 & 18.7 & -3.7 & 5.8
\\ 

(0 $\bar{4}$ 0) & -3.6 & 9.8 & -3.7 & -1.0 & (0 $\bar{8}$ 0) & -2.2 & 21.5 & -3.7 & -2.4
\\
 




(0 2 2) & -3.8 & -1.5 & -4.3 & -1.8 & (0 4 4) & -3.5 & -5.7 & -3.6 & -9.7
\\ 

(0 $\bar{2}$ 2) & -3.9 & 5.0 & -4.4 & 5.6 & (0 $\bar{4}$ 4) & -3.6 &  9.1 & -3.8 & 13.1
\\

(2 0 2) & -3.6 & 2.2 & -4.4 & 2.0 & (4 0 4) & -3.9 & 3.8 & -4.7 & 3.2     
\\ 




($\bar{1}$ $\bar{1}$ $\bar{1}$) & -3.8 & 2.8 & -0.1 & 0.4 & (3 3 3) & -3.6 & -5.6 & -0.04 & -0.7
\\ 

(1 1 $\bar{1}$) & -3.8 & 0.7 & -0.1 & -0.3 & (3 $\bar{3}$ 3) & -3.7 & 9.1 & -0.06 & 0.8
 \\ 



\hline 
\hline 
\end{tabular}
} 
\end{table}
\end{center}

\vskip -0.9cm

In Fig 2, we show the contact hyperfine frequencies as a function of E-field for 6 different groups (shells) of sites around the donor. Fig 2a and 2b are for sites along [100] with distances of $a_0$ and $2a_0$ from the donor respectively. The degenerate point at $E=0$ splits into three curves for these groups, as discussed before. Fig 2c and 2d are for two shells along [110], while 2e and 2f are for those along [111]. The 12 equivalent points in a shell of [110] split into 3 groups, while the 4 equivalent points of a shell of [111] split into two groups.  

As shown in Fig 2, the frequency axes of the various plots do not overlap with each other even at fields of 3 MV/m. This means that the shells can be distinguished even in the presence of electric fields. To show the detectability of the points within a shell, we have incorporated an uncertainty of 0.1 MV/m (estimated from Fig 1 of Ref \cite{Bradbury.prl.2006}) in the E-field, and represented by error bars in Fig 2. This shows that even with some inhomogeneity in the field, most sites can still be identified. Furthermore, the distinguishability improves at higher E-fields, and for shells farther away from the donor within the Bohr-radii due to their increased splitting. The shell comprising of the nearest neighbor sites of the donor (Fig 2e) may not be distinguishable due its small splitting and low resonance frequencies. A measurement of the shift of the contact hyperfine frequencies directly provides a measure of the shift in the electron probability density as $\beta(E)-\beta(0) \propto |\psi(E)|^2-|\psi(0)|^2$. 

In Fig 2g and 2h, we show the Stark shift of one of the dipolar components $B_{zz}$. These terms are considerably small in magnitude relative to $\beta$. However, these give a measure of the field induced deformation in the wavefunction about the donor nucleus through the dipole operator (eq (1)). Theoretical models involving donor potential and crystal band-structure may be optimized to fit these hyperfine components, and thus to improve their accuracy.

We provide quantitative Stark shift data in Table 1 for the components $\beta$ and $B_{zz}$ of the six shells considered in this work. To concisely present the Stark shifted hyperfine frequencies, we fitted the curves of Fig 2 to the form ${\Delta}\alpha(\vec{E}) = \alpha(0)(\eta_{2}{E}^{2} + \eta_{1}E)$, 
where $\alpha=(\beta,B_{ij})$, $\eta_2$ and $\eta_1$ are the quadratic and the linear Stark coefficients respectively. The values of $\eta_2$ and $\eta_1$ obtained from the fit are listed in Table 1. Given the hyperfine tensor component at $E=0$ for a shell and an applied field value, one can calculate both the Fermi contact hyperfine coupling and a dipolar tensor component using this table. As an example, the zero-field hyperfine frequencies of $\beta$ and $B_{zz}$, reported in Table II of Ref \cite{Hale.prb.1969} for the first shell along [100], are 2981 and 41.4 kHz respectively. With $E=4$ MV/m, and using Table I and the qudratic eq, we predict that $\beta$ and $B_{zz}$ of the site $(0,4,0)$ decrease by 243 and 3.3 kHz respectively, a net change which should be experimentally detectable. In comparison, the other two non-equivalent sites, $(0, \bar{4},0)$ and $(0,0,4)$, are shifted by 55 and 160 kHz in $\beta$ and 0.7 and 2.2 kHz in $B_{zz}$ respectively. Therefore, these 3 sites should be distinguishable.

\begin{figure}[htbp]
\center
\epsfxsize=3.0in
\epsfbox{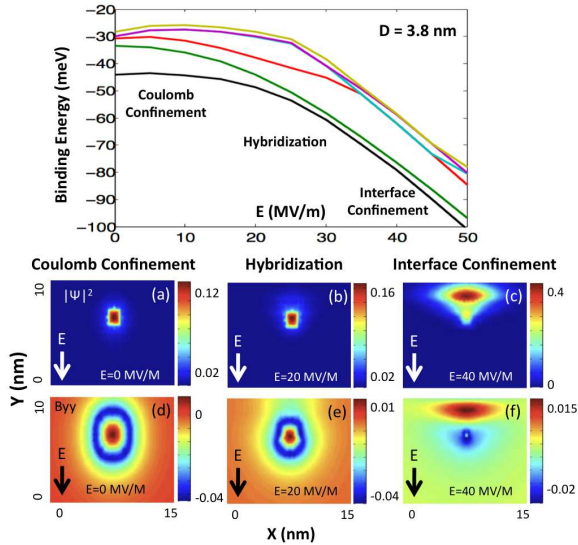}
 \caption{Top row: The stark shifted spectrum of a P donor at 3.8 nm from an oxide interface. Bottom row: (a), (b) and (c) show the P donor ground state wavefunction at three different E-fields, while (d), (e) and (f) show the corresponding hyperfine maps in the form of $B_{yy}$ tensor component. The E-field is perpendicular to the interface.}  
\end{figure}

Lastly, we give an illustration of how the method is useful for understanding the extent of wavefunction deformation and quantum confinement of direct relevance to quantum nanoelectronics \cite{Rogge.NaturePhysics.2008}. Fig. 3 shows the spectrum and the wavefunctions of a P donor at 3.8 nm depth from the oxide interface subjected to electric fields 0, 20 and 40 MV/m (Fig 3a, 3b, and 3c respectively). In this regime, the donor wavefunction can be modified adiabatically by the field {\cite{Calderon.prl.2004, Smit.prb.2003, Martins.prb.2004, Wellard.prb.2005}}, as the electron makes a transition from a purely Coulomb confined state at $E=0$ to a purely 2D confined state at the interface at $E=40$ MV/m. In the intermediate field regime ($E=20$ MV/m), the electron resides in a superposition of Coulomb bound and surface bound states (Fig 3b). This serves an example of controlled wavefunction engineering by electric fields. An associated  dipolar tensor component, $B_{yy}$ for example, is shown on the 2nd row of Fig 3 (d, e and f), and reflects the gradual symmetry change of the donor wavefunction.

In conclusion, we proposed the measurement of hyperfine maps of donors as a means of experimentally characterizing field induced distortions and symmetry changes of electron wavefunction. The nuclear spin of a ${}^{29} \textrm{Si}$ atom can act as a probe of the donor wavefunction, providing a site by site map of electron localization. Such maps can help us investigate the unknown electronic wavefunctions in novel Si nanostructures for a host of quantum nanoelectronic applications, and fine tune various modeling techniques at the atomic scale. The predictions of the Stark shift of the hyperfine tensors for six different shells near a P donor indicate that experimental detection of engineered wavefunctions is feasible for lattice sites in the immediate vicinity of the donor, thus providing a probe of the wavefunction at sub-Bohr radius resolution. 

This work was supported by the Australian Research Council, NSA and ARO under contract number W911NF-08-1-0527. NEMO-3D was initially developed at JPL, Caltech under a contract with NASA. NCN/nanohub.org computational resources were used. 

Electronic address: park43@purdue.edu

\end{document}